\documentclass[12pt]{article}

\usepackage{scicite}
\usepackage{times}
\usepackage{tabularx}
    \usepackage[hyphens,spaces,obeyspaces]{url}
\usepackage{color}
\usepackage{subcaption}
\usepackage{caption}
\usepackage{graphicx}
\usepackage{mdwlist}

\usepackage{multicol}
\usepackage{multirow}
\usepackage{ctable}
\usepackage{lineno}
\usepackage{lipsum}
\usepackage[headheight=0pt,headsep=0pt]{geometry}
\usepackage{authblk}

\usepackage[citestyle=numeric, bibstyle=numeric, natbib=true, backend=bibtex]{biblatex}
\graphicspath{{figures/}}       

\usepackage{xcolor}
\newcommand\xxx[2][Name]{}

\newenvironment{sciabstract}{%
\begin{quote} \bf}
{\end{quote}}

\newcounter{lastnote}

\newcommand*\dash{\unskip\kern.16667em---\penalty\exhyphenpenalty
        \hskip.16667em\relax
}

\newcommand\blfootnote[1]{%
  \begingroup
  \renewcommand\thefootnote{}\footnote{#1}%
  \addtocounter{footnote}{-1}%
  \endgroup
}

\title{\LARGE{Computer Security Risks of Distant Relative Matching in Consumer Genetic Databases}}

\author[1,3,4,5]{Peter M. Ney}
\author[2,4,5]{Luis Ceze}
\author[1,3,4,5]{Tadayoshi Kohno}
\affil[1]{\small{Security and Privacy Research Lab}}
\affil[2]{\small{Molecular Information Systems Laboratory}}
\affil[3]{\small{Tech Policy Lab}}
\affil[4]{\small{Paul G. Allen School of Computer Science \& Engineering}}
\affil[5]{\small{University of Washington}}

\date{}

\begin{document}

\maketitle

\begin{sciabstract}
Consumer \blfootnote{*corresponding e-mail: dnasec@cs.washington.edu}genetic testing has become immensely popular in recent years and has lead to the creation of large scale genetic databases containing millions of dense autosomal genotype profiles. One of the most used features offered by genetic databases is the ability to find distant relatives using a technique called relative matching (or DNA matching). Recently, novel uses of relative matching were discovered that combined matching results with genealogical information to solve criminal cold cases. New estimates suggest that relative matching, combined with simple demographic information, could be used to re-identify a significant percentage of US Caucasian individuals. In this work we attempt to systematize computer security and privacy risks from relative matching and describe new security problems that can occur if an attacker uploads manipulated or forged genetic profiles. For example, forged profiles can be used by criminals to misdirect investigations, con-artists to defraud victims, or political operatives to blackmail opponents. We discuss solutions to mitigate these threats, including existing proposals to use digital signatures, and encourage the consumer genetics community to consider the broader security implications of relative matching now that it is becoming so prominent.
\end{sciabstract}

\vspace*{10pt}

\nolinenumbers
\baselineskip20pt

\section*{Introduction}

\vspace*{-5pt}

Direct-to-consumer (DTC) genetic testing services have become immensely popular in recent years~\cite{mit-tech-year-testing}. For under \$100, DTC services will genotype customers using high-density autosomal SNP microarrays that probe between 500,000-1,000,000 markers~\cite{23andMe, ancestry, my-heritage}. These genetic profiles are analyzed to help customers better understand their ancestry and health, or for general curiosity. DTC companies also allow customers to download their raw genetic profiles so they can analyze the data themselves or upload profiles to third-party services that can aggregate profiles from different DTC companies and offer advanced features. Many third-party service providers were created for amateur or academic purposes, but recently, they have become quite large, in some cases containing close to 1 million profiles~\cite{bloomberg-tech}. We distinguish between DTC companies, which produce genetic data directly from biological material, and third-party services that only aggregate and analyze genetic data produced from other sources.

One of the most popular features offered by both DTC and third-party services is long range familial relative matching (or DNA matching). With high-density genotype data it is possible to accurately identify distant relatives, including relatives as distant as 3rd cousins or further, in large databases that include millions of profiles~\cite{mountain2012}. Relative matching algorithms work by finding large DNA segments that are shared between individuals, known as identical by descent (IBD) segments, that are indicative of shared ancestry because they have been inherited from a recent common ancestor without intervening recombination. The total quantity and distribution IBD segments between two individuals is then used to predict their familial relationship~\cite{gusev2009whole}. Analysis with MyAncestry data showed that matching relatives are quite common, with over 44\% of profiles returning a third cousin and 10\% with a second cousin or closer~\cite{Erlich350231}. We can expect that the probability of a match will only increase as these databases grow larger.

Relative matches can be profound and sometimes life changing because of the cultural and legal significance of family relationships. For example, matches routinely reunite adopted children with their birth parents and identify cases of unexpected paternity~\cite{dna-success-stories,isogg-success,parents-gift-divorce}. However, it was the high profile use of relative matching by law enforcement to identify the suspected Golden State Killer that has perhaps drawn the most attention. To solve this case, law enforcement used DNA samples obtained from old crime scenes to construct a valid DTC genetic profile; this profile was uploaded to GEDMatch, a third-party genetic ancestry service that supports relative matching~\cite{washingtonpost-gsk,nytimes-gsk}. Investigators identified third cousins of the suspect, which enabled them to narrow their search to the suspected killer~\cite{washingtonpost-gsk}. Since the Golden State Killer case, genetic genealogy approaches have been used over a dozen times to solve other cold cases and identify deceased individuals~\cite{forensic-magazine,dna-doe-project}. Business like Parabon Nanolabs labs now offer genetic genealogy services to law enforcement to help them solve cases using this technique~\cite{parabon-nanolabs}.

Relative matching in consumer genetic databases significantly enhances the capabilities of law enforcement to match distant relatives when compared to government maintained forensic databases, like the FBI maintained CODIS database. CODIS primarily maintains alleles for a small number of 13-20 core, short tandem repeat (STR) markers~\cite{budowle1998codis,fbi-faq}. Relative searching with STR markers relies on comparing markers directly, which limits the CODIS database to finding closely related family members~\cite{ge2011comparisons}. Consumer databases also contain a number of other advantages over government forensic databases. They contain profiles from a more diverse population because they include data from the general public, not just suspected or convicted criminals, and relative matching in consumer databases avoids regulations that restrict or preclude relative matching in government databases~\cite{ram2018genealogy,science-progress-transparency}. Therefore, it is no surprise to see law enforcement rapidly adopt relative matching techniques on consumer databases to solve crimes.

The growing prominence of DTC and third-party consumer genetic databases in important applications, like criminal investigations, raises significant computer security (a.k.a., cyber security) concerns. In effect, third-party genetic databases designed for amateur ancestry analysis are now used as criminal forensic databases \dash except there do not appear to be the same rigorous computer security standards that exist in government maintained forensic databases. Most concerning is that other individuals can leverage the same genetic genealogy techniques as law enforcement: malicious actors can re-identify individuals, spoof profile metadata (e.g., identity of the profile), or even make entirely fake profiles by directly manipulating the digital genotype data. As these databases continue to be used beyond their original intended purpose, criminals or other malicious actors may be motivated to compromise or manipulate databases that support relative matching. In this work we describe potential risks associated with relative matching in consumer genomic databases and discuss what services can do to enhance their safety.

\paragraph{Next steps.}
We argue that given the high profile use of these databases, now is the time to have an industry-wide conversation about the computer security risks relevant to DTC and third-party consumer genetic databases. Since the issues we raise have the potential to be industry-wide, and since many of the observations we make are natural extensions of prior observations (but with a computer security perspective), we would consider it improper to disclose our findings to only the companies and actors in this field that we can enumerate 
\dash doing so would only give them and not others the ability to proactively consider mitigations. By disclosing our findings of potential risks in this work, as well as a further discussion of defenses, we hope to catalyze the incorporation of strong defenses in \emph{all} DTC and third-party consumer databases, and we hope to engender the appropriate level of caution by \emph{all} parties (such as those from law enforcement) who might seek to use present-day DTC and third-party consumer databases in their activities.

\vspace*{-5pt}

\section*{Adversarial Goals}

\vspace*{-5pt}

Support for unrestricted relative matching exposes consumer genetic databases to additional computer security and privacy related risks than what might be anticipated. Other issues, like data theft or discriminatory use of genetic data, are certainly a concern, but in this work we emphasis how relative matching, in particular, creates new problems. Consider the wide range of malicious actors, with varying capabilities, that have motivation to take advantage of relative matching services. This includes criminals with unidentified forensic DNA evidence that want to remain anonymous, con-artists that could use fabricated relatedness to gain the trust of their victims, or even sophisticated political or intelligence actors that want to blackmail or harm the reputation of their opponents (see Table~\ref{tbl:actors} for an extended list of actors and their goals). We categorize the goals of adversaries into three groups:

\vspace*{-5pt}

\begin{enumerate}
  \item \textbf{Avoid identification.} The goal is to remain anonymous or difficult to detect with relative matching queries. This is precisely the situation where a criminal perpetrated a crime and left physical DNA evidence, like the Golden State Killer, and is trying to avoid detection by law enforcement using ancestry analysis. This is also useful for individuals, like known fugitives, that have assumed a false identity to remain undetected \dash in fact, just recently, relative matching on GEDMatch was used to identify an individual that had assumed the identity of a deceased 8-year old boy for over 20 years~\cite{fugitive-wp}.
  \item \textbf{Determine a target's identity and relatives.} It has already been shown that relative matching can be used to re-identify a large number of US individuals, and just as law enforcement has used relative matching to identify unknown individuals, malicious actors could do the same~\cite{Erlich350231}. Such re-identification could threaten the privacy of anonymized research subjects or endanger foreign operatives and undercover agents that have assumed a false identity for legitimate purposes. Matching results could also be used to uncover sensitive details, like hidden children. For example, if a high profile individual, like a politician or celebrity, had an unknown or hidden child then an attacker could use this knowledge to discredit or blackmail that individual.
  \item \textbf{Remove or forge new relatives.} Criminals could remove or use forged relationships for a number of purposes: con-artists could forge relations with a target to gain their trust, perpetrators could put blame on others by forging profile data, and parents could use falsified profiles to misattribute paternity to another person.
\end{enumerate}

\begin{table}[t!]
\begin{center}
 \resizebox{\textwidth}{!}{%
 \begin{tabular}{| l | l |} 
 \hline
 \textbf{Malicious Actor} & \textbf{Motivations} \\
 \hline\hline
  \multirow{2}{*}{Perpetrator of Unsolved Crime or Fugitive} & Avoid Identification \\
  & Receive Advanced Warning of Pending Discovery \\
  \hline
  Confidence Trickster & Defraud Target \\
  \hline
  Paparazzi or Stalker & Reveal Target's Ancestry \\
 \hline
  \multirow{3}{*}{Unwilling Parent} & Avoid Paternity Obligations \\
  & Imply False Paternity on a Target \\
  & Avoid Discovery of Infidelity \\
  \hline
  Genetic Surveillance & Link Unknown Genetic Profile to Identity \\
  \hline
  Curious Individual & Re-Identify Anonymous Public Research Profiles \\
  \hline
  \multirow{2}{*}{Political Operative} & Damage Opponent's Reputation \\
  & Blackmail \\
  \hline
  \multirow{2}{*}{Intelligence Agency} & Identify Foreign Operatives \\
  & Blackmail \\
  \hline
 \end{tabular}}
\end{center}
    \caption{Possible malicious actors and their motivations to attack genetic databases that support relative matching.}
\label{tbl:actors}
\end{table}

\section*{Threat Vectors}

\vspace*{-5pt}

\begin{figure*}[t!]
\includegraphics[width=\linewidth]{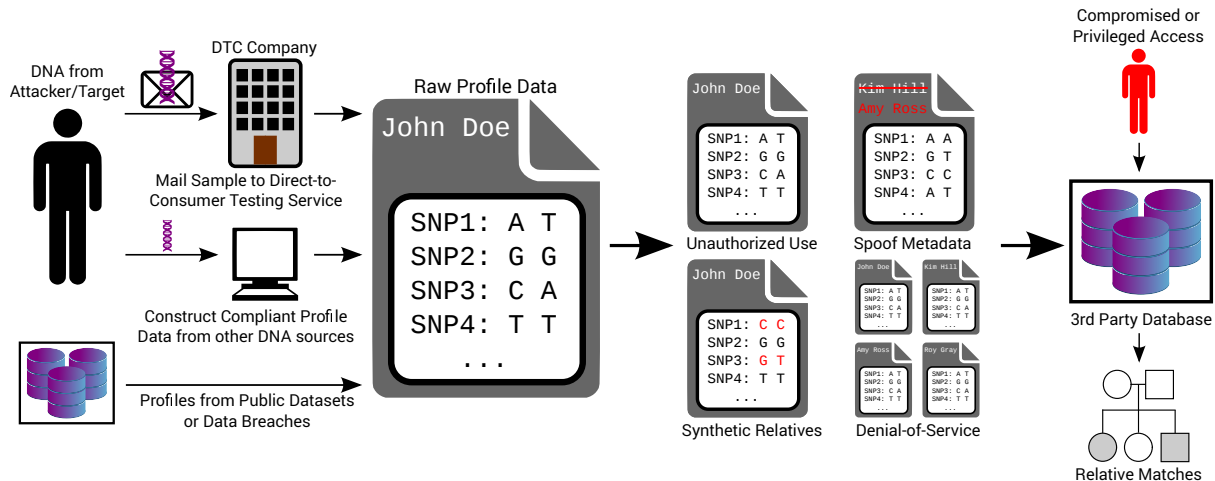}
\caption{Overview of threats to 3rd-party genetic databases that support relative matching. DNA profiles can be obtained in a number of ways, including through standard DTC genetic testing, by indirectly generating them from other DNA sources, or from publicly available data. An attacker can take advantage of relative matching by uploading manipulated profile data, or if they have direct access, by manipulating the database itself.}
\label{fig:attack_vectors}
\end{figure*}

The open nature of consumer genomics, especially third-party services, gives attackers many ways to manipulate relative searches. Below we describe different methods that a malicious actor can use to achieve different objectives: unauthorized use of DNA data or profiles, metadata spoofing, synthetic relatives, denial-of-service, and directly deleting or modifying existing genetic profiles (Table~\ref{tbl:attack-vectors}; Figure~\ref{fig:attack_vectors}).

\paragraph{Unauthorized Use of DNA Profiles.}

This is when someone fraudulently gathers or uses a target's genetic profile in a relative matching query without authorization. This would be useful to anyone wanting to determine an individual's identity or familial relationships. To use a target's genetic profile in a relative matching query, the attacker needs either access to the digital genotype data or be able to collect a DNA sample from the target to generate the profile. In some cases, the attacker will already have access to the digital profile of a target. For example, there are many public research datasets that contain high-density SNP microarray or whole genome sequencing data for many anonymous research subjects~\cite{10002015global}. This data could easily be converted into compliant DTC genetic profiles that could be uploaded to third-party databases. Moving forward, data breaches may be another source of genetic profiles. There have already been security issues at genetic testing companies. For example, in June 2018 it was discovered that over 92 million user passwords and email addresses from a major DTC company were leaked~\cite{my-heritage-incident}.

When the attacker does not have the target's genetic profile data but has physical access to them, a competent attacker can collect a physical sample from the target to construct a genetic profile. Currently, major DTC companies require saliva or buccal samples to generate profiles, which may be difficult to obtain from a target. However, there are already examples of amateurs collecting and processing DNA from heterogeneous samples. In 2013, the artist Heather Dewey-Hagborg, in her project Stranger Visions, collected random DNA around New York City from sources like cigarette butts and gum and analyzed this DNA to generate face portraits using forensic DNA phenotyping algorithms~\cite{heather-stranger-visions}. In a similar fashion, an attacker could use DIY-bio spaces, available in many cities, to extract DNA and sequence it using a third-party sequencing provider. Heather Dewey-Hagborg also demonstrated that it is easy to obtain anonymous saliva samples when she ordered some over the Internet using only an academic email and mailing address and then sent the saliva to a DTC company for processing~\cite{heather-t3511}. Some actors, like nation states, could generate profiles from DNA samples directly using their own facilities.

\begin{table}[t!]
\begin{center}
 \resizebox{\textwidth}{!}{%
 \begin{tabular}{| l | l | l |} 
 \hline
 \textbf{Attack} & \textbf{Objective} & \textbf{Expertise} \\
 \hline\hline
  \multirow{2}{*}{Unauthorized Use of DNA Profiles} & Uncover Identity of Target & \multirow{2}{*}{Low} \\
  & Determine Target's Relatives & \\
  \hline
  \multirow{2}{*}{Spoof Metadata} & Hide Identity from Relative Searches & \multirow{2}{*}{Low} \\
  & Implicate Wrong Individual & \\
  \hline
  \multirow{2}{*}{Synthetic Relatives} & Misdirect Relative Searches & \multirow{2}{*}{Medium} \\
  & Create False Relatives of a Target & \\
  \hline
  Denial-of-Service & Thwart Relative Matching & Medium \\
 \hline
  \multirow{2}{*}{Compromised or Privileged Access} & Add, Delete, or Modify Relative Matches & \multirow{2}{*}{High} \\
  & Corrupt Audit Log & \\
  \hline
 \end{tabular}}
\end{center}
    \caption{Possible attacks against genetic databases with relative matching, the objective of these attacks, and necessary expertise required to execute a given attack.}
\label{tbl:attack-vectors}
\end{table}

\paragraph{Spoof Metadata.}

One of the simplest things a malicious actor can do is to falsify metadata \dash the information associated with the genetic profiles like name or email address \dash when a DNA sample or raw profile data is sent to a DTC or third-party service. In some cases, this metadata is the only information linking the genetic profile to a particular individual.

This is clearly useful to someone trying to avoid identification. For example, a criminal could upload their own profile data, which they could easily obtain from a DTC service, under a false identity. This could confuse or slow down an investigation because the criminal's profile would appear to be another individual. It is likely that in criminal investigations any suspects would be re-tested by law enforcement, which would reveal that the metadata was falsified. However, the criminal could make this more difficult by spoofing the identity of an individual that is difficult to reach, like someone that is deceased, living in another country, or is entirely fictitious. If the criminal uploaded the data of another individual under their own identity they might avoid suspicion in cases where law enforcement had more direct access to a genetic database to run queries.

\paragraph{Synthetic Relationships.}

The raw genotype data itself, including SNPs, can also be modified. This is a problem because some third-party services do not confirm that uploaded profile data was produced by a reputable DTC service and left unmodified. IBD segment sharing has been simulated in prior work to evaluate IBD detection algorithms, and we believe that similar techniques can be used to generate forged profiles that are designed to look like relatives a given individual, which we call \textit{synthetic relatives}~\cite{gusev2009whole,mountain2012}. If synthetic relative profiles were then uploaded to third-party services, those profiles could trick relative matching algorithms into making spurious relative matches.

The ability to create false relatives can lead to a number of security problems. Consider the case of a perpetrator that is trying to avoid identification. The simplest case is when the perpetrator has no identifiable relatives in a third-party database; they could upload a distant synthetic relative under a false identity to misdirect a genetic genealogy investigation. This may also give the perpetrator advanced notice of a pending investigation because the person falsely associated with the synthetic relative may be contacted for questioning about their family history and may have an existing relationship with the perpetrator.

\begin{figure*}[t!]
\includegraphics[width=\linewidth]{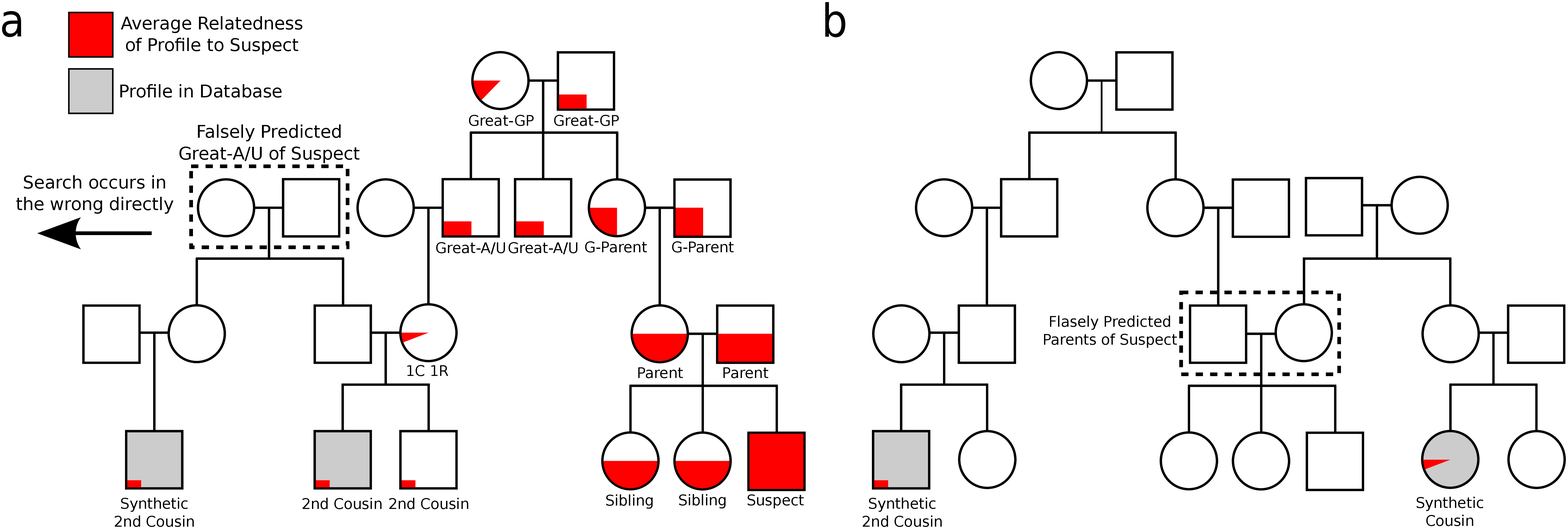}
\caption{Examples of attacks using synthetic relatives. \textbf{a}, A criminal suspect wants to avoid identification when their 2nd cousin is already in a database. The suspect uploads a synthetic second cousin using the identity of an individual related to the 2nd cousin but not the suspect. This makes it appear that the suspect is on the wrong branch of the family tree. \textbf{b}, The suspect uploads two synthetic relatives, on different family branches, to falsely implicate a couple as parents of the suspect.
\label{fig:synthetic_attacks}}
\end{figure*}

A more challenging scenario is when the perpetrator already has distant relative profiles in third-party databases (which the perpetrator could identify using their own profile or a partially falsified one). In this case, the perpetrator could upload synthetic relatives, but with the identity of someone on a distant branch of the family tree that is not related to the perpetrator but is related to a real relative already in the database (see Figure \ref{fig:synthetic_attacks}a for a visual example). Whether this works in general will depend on how many matching relatives already exist in the database, the overall topology of the family tree, and the existence of other sources of corroborating information, like known family trees. Distant synthetic relatives can also be used to falsely implicate a particular target. For example, in some situations, two matching relatives on different branches of a family tree can narrow a search to a small number of individuals (Figure \ref{fig:synthetic_attacks}b). Any falsified, synthetic relatives would be discovered if the identity associated with them was ever closely investigated and re-tested. However, since synthetic matches are designed to appear like relatives of the perpetrator and not the perpetrator themselves, it is possible that the synthetic relatives would not be scrutinized.

It is not just perpetrators but other types of actors that could use synthetic relatives maliciously. For example, a con-artist could upload a synthetic profile designed to look like the child of a target, using the con-artist's name, to falsify a parent-child relationship. Politically motivated actors could also use synthetic children to imply a child born out of wedlock to tarnish their opponent's reputation. In these cases, the relationship might seem serendipitous and not be re-tested (or not be re-tested until after significant damage has been done).

We propose a simple approach to generate synthetic relatives by splicing SNPs from multiple profiles together. To begin, the attacker starts with the genetic profile of the target that will be related to the synthetic relative. Then using another profile as a base \dash say from a public database or an entirely synthetic profile \dash the attacker stitches in segments from the target profile into the base profile based on the length and number of IBD segments that would be expected from the desired relationship (e.g., parent, 2nd cousin, etc.). The segment regions could be randomly selected or copied from real relative matches with the desired relationship (which are commonly published on amateur ancestry blogs).

\begin{figure*}[t!]
\includegraphics[width=\linewidth]{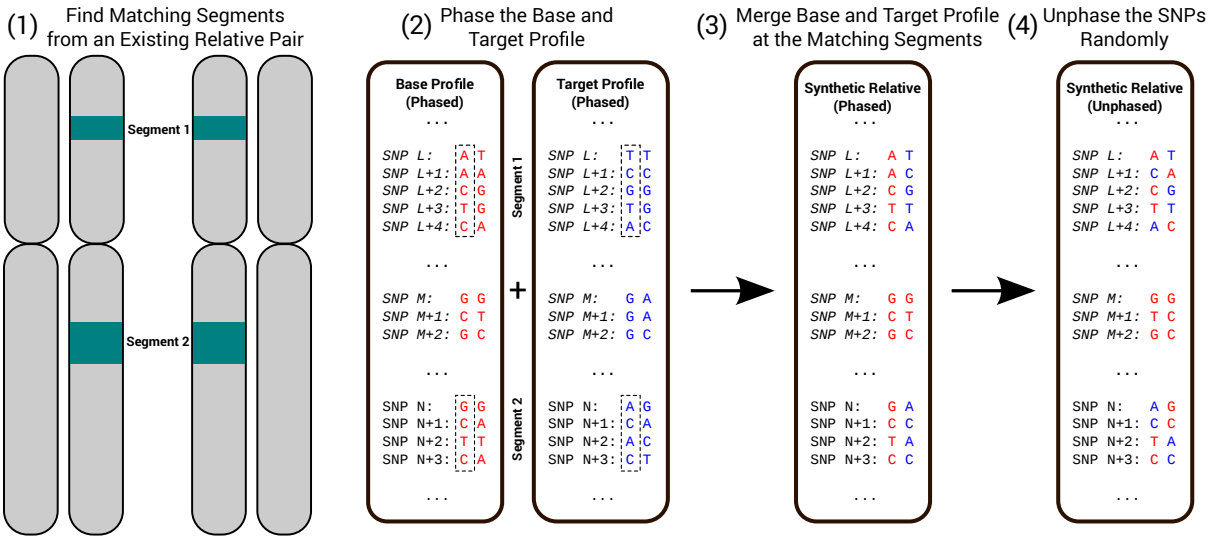}
\caption{Proposed method to generate synthetic relatives.}
\label{fig:synthetic_relative_construction}
\end{figure*}

Segments could be spliced from the target profile into the base profile by naively copying over both alleles in each segment region. However, this would make the matching segments appear to be full-IBD (i.e., matching on both chromosomes), which would only be expected for close relatives, like siblings. To construct half-IBD segments, only SNPs from one phase should be spliced into the base profile~(Figure \ref{fig:synthetic_relative_construction}). This technique could be extended to make a synthetic profile related to many individuals by splicing additional target profiles together in a similar manner. It is important to note that the synthetic relative will appear related to the base profile as well, which might cause undesirable matches if a real genetic profiles was used as the base profile. To summarize, synthetic relative construction could work as follows:

\begin{enumerate}
  \itemsep0em 
  \item Determine the segment regions to copy from a an existing relative pair.
  \item Phase the genotypes of the target and base profiles.
  \item For each segment region, copy over the SNPs from one phase of the target profile and replace those SNPs in one phase of the base profile.
  \item Unphase the new modified base profile by randomly shuffling the SNPs. This modified base profile is the final synthetic relative profile.
\end{enumerate}

\paragraph{Denial-of-Service.}

There is no limit on the number of accounts and profiles that an attacker can create and upload if account information is not verified. This can be useful to an attacker that is trying to hide their identity because they can upload synthetic relatives under many different accounts. This will cause an excessive number of relative matches that could overwhelm any real matches with fake ones. However, this technique might arouse suspicious because the number of matches could appear unrealistic. Another approach is to spoof a person with a large family or who is a descendent of someone with a large number of children, like a sperm donor. This could make it more difficult to narrow down the search to a manageable list of suspects because the extended family would be very large.

\paragraph{Compromised or Privileged Database Access.}

The most powerful technique to manipulate relative matches is when an attacker has compromised or privileged access to a genetic database. With privileged access, an attacker is able to add, remove, or modify any profile in the database to alter relative matches or modify logs used for auditing. This threat is particularly acute when services have poor security practices. Here we note that experience in the computer security community suggests that non-commercial services, especially those run by a small number of people, can often have poor security practices; further, industries that have not yet received adversarial pressure often do not implement state-of-the-art security best practices. This threat is also relevant to actors, like governments, that are technically capable or to insiders that already have high levels of access.

\section*{Discussion}

Long range familial matching is not new, but it was not until the recent explosion of high-density consumer genetic testing that it became so effective and consequential. The ability to match distant genetic relatives at a population scale is profound and brings a new set of computer security challenges. The recent appropriation of amateur ancestry databases as an investigative tool for law enforcement significantly raises the stakes.

In this work we described a wide range of actors with varying capabilities that have interest in modifying or corrupting relative matching queries, which makes the threat space especially complex. There are a number ways to minimize the risk of relative matching attacks. A major issue is that some third-party services do not confirm that all profiles originate from legitimate DTC services. A failure to authenticate profiles  may allow attackers to upload digitally constructed or altered DTC genetic profile data files. This attack vector can be eliminated if all raw profiles are known to be generated by a reputable DTC testing service. This can be accomplished if the raw data files are digitally signed using a cryptographic key controlled by the DTC company, as was recently suggested by Erlich et.\ al.,\ or by exchanging data using APIs instead of letting users directly upload their raw genetic profiles~\cite{Erlich350231}. However, we suggest going a step further by having the genotyping instrument itself sign and attest to the data it generates, including instrument id, date, and time of data generation. As a side-effect, digitally signed genetic profiles will prevent law enforcement from uploading raw profiles in cases where profiles cannot be generated directly from a DTC service and will instead require law enforcement to work with DTC companies directly.

Digital signatures will ensure that the genotype data was created by a DTC testing service, but will not confirm that metadata associated with a profile is legitimate. To do this DTC and third-party companies should confirm the identity of their customers. It is a common practice for people to upload profiles on behalf of their relatives, so it may not be possible to verify the identity of specific profiles that are generated; however, at a minimum, the identity of the account owner should be verified. Identity verification also minimizes the risk of denial-of-service attacks because it naturally limits the number of account that can be created. Account identity could also be further tied to the profile data via digital signatures.

Finally, we have concerns about the general computer security practices of consumer genetic databases, especially third-party services that have not previously faced significant adversarial pressure. Given the sensitivity, size, and way third-party services are being used, we argue that it is important that they adopt software security best practices. This can include instituting better user authentication procedures, routine auditing and penetration testing, and frequent software updates, among others.

In the future, the efficacy of different relative matching attacks will depend largely on the population coverage in these database and the presence of other sources of information like family histories. The more dense these databases, the more difficult it will be to avoid identification. However, other attacks, like re-identification, will become correspondingly easier. In any case, whenever relative matches are used in important applications (e.g., criminal investigation, paternity, etc.) all results should be corroborated with additional testing.

\paragraph{Acknowledgements}\mbox{}\\
This research was supported in part by the University of Washington Tech Policy Lab and the Torode Family Professorship.

\paragraph{Competing Interests}\mbox{}\\
The authors declare no competing interests.

\printbibliography

\end{document}